\documentclass{aa}
\usepackage{times}
\usepackage{graphics}
\usepackage{psfig}

\begin{document}

\newcommand{\kms}   {\mbox{km.s$^{-1}$}} 
\newcommand{\Teff}  {\mbox{T$_\mathrm{eff}$\,}}
\newcommand{\FeH}   {\mbox{[Fe/H]}\,}
\newcommand{\logg}  {\mbox{$\log$ g\,}}
\newcommand{\grav}  {\mbox{\logg}}
\newcommand{\masy}  {\mathrm{\,mas\,y}^{-1}}
\newcommand{\Mv}    {\mbox{M$_\mathrm{V}$\,}}
\newcommand{\Blim}  {B_\mathrm{lim}}
\newcommand{\Ha}    {H$_\mathrm{\alpha}$}
\newcommand{\Hb}    {H$_\mathrm{\beta}$}
\newcommand{\vr}    {v$_\mathrm{r}$}
\newcommand{\pc}    {\,\mathrm{pc}}
\newcommand{\kpc}   {\,\mathrm{kpc}}
\newcommand{\K}     {\,\mathrm{K}}
\newcommand{\vsini} {v$\sin$ i\,}
\newcommand{\tgm}   {\mbox{(\Teff, \logg, [Fe/H])\,}}

\thesaurus{04                % A\&A Section 04 : Astronomical data bases
		 (04.01.2;   % Atlases
		  08.01.1;   % Stars: abundances,
		  08.01.3;   % Stars: atmospheres,
		  08.06.3;   % Stars: fundamental parameters,
                  10.19.2)}   % Galaxies: stellar content

\title{A database of high and medium-resolution stellar spectra
\thanks{based on observations made on the 193cm telescope at 
the Haute-Provence Observatory, France.}
      }

\author{Ph. Prugniel \inst{1} \and C. Soubiran \inst{2}}

\offprints{Ph. Prugniel (prugniel@obs.univ-lyon1.fr)}

\institute{CRAL-Observatoire de Lyon, CNRS UMR 142, F-69561 Saint-Genis Laval, France
\and 
Observatoire de Bordeaux, CNRS UMR 5804, BP 89, F-33270 Floirac, France}

\date{Received ; accepted }

\titlerunning{Database of stellar spectra}
\maketitle

\begin{abstract}
 We present a database of 908 spectra of 709 stars obtained with the 
ELODIE spectrograph at Observatoire de Haute-Provence.
52 orders of the echelle spectra have been carefully fitted together to provide
continuous, high-resolution spectra  in the 
wavelength range \, $\lambda\lambda = 410 - 680 $~nm. The archive provides a large coverage of the 
space of atmospheric parameters :  
\Teff from 3700 K to 13600 K, \logg \, from 0.03 to 5.86 
and  \FeH from -2.8 to +0.7. 
At the nominal resolution, R=42000, the mean signal-to-noise ratio is
150 per pixel. The spectra given at this resolution are normalized to their 
pseudo-continuum and are intended to serve for abundance studies,  spectral
classification and tests of stellar atmosphere models.
A lower resolution version of the archive, at R=10000, is calibrated in 
physical flux with a broad-band photometric precision of 2.5\% and
narrow-band precision of 0.5\%. It is scoped for stellar population 
synthesis of galaxies and clusters, and for kinematical investigations of
stellar systems.

The archive is distributed in FITS format through the HYPERCAT and CDS  databases. 


\end{abstract}

\section{Introduction}

Spectral libraries covering the HR diagram with medium to high resolution and large 
spectral range are essential tools in astronomy. We have identified two areas where 
there will be a special need of such libraries in the coming years : the automated 
parameterization of stellar spectra and  the spectral 
synthesis of stellar populations of galaxies. New multi-object 
spectrographs on large telescopes,
like Giraffe on the VLT for example, will soon make it possible to observe hundreds of objects during
the same exposure with an unprecedented resolution and  spectral coverage. The challenge
will be to extract the maximal information in a reasonable time. 

 For stars, medium to high resolution spectra are the basis for determining
radial velocities, 
atmospheric parameters \tgm and eventually projected rotational 
velocities, \vsini. 
 A new method, called TGMET, is emerging to make the parameterization 
of stellar spectra in terms of temperature, gravity and metallicity fast and 
automatic (Katz et al \cite{k98}). 
It relies on the direct comparison of a target spectrum to a library of 
reference spectra with  known atmospheric
parameters (Soubiran et al \cite{s98}). TGMET was originally dedicated to the 
measurement of F, G, K
stars observed with ELODIE. The present paper, which
 continues the TGMET project, participates to release these limitations. The two major
points addressed are (1) the extension of the reference library to all spectroscopic
types and luminosity classes and its densification, and (2) to resample the spectra in
wavelength (i.e. remove the instrumental signature in ELODIE spectra and connect the
echelle orders together) in order to enable comparison of spectra from any origin.

The TGMET project appeared to be also very useful for extragalactic studies where
medium resolution libraries with large coverage in stellar parameters are needed.
Stellar libraries, calibrated in flux, are used to
model composite spectra through stellar population synthesis (eg. the PEGASE program,
 Fioc \& Rocca-Volmerange \cite{f97}).
An important issue is to disentangle the effects of age and 
metallicity and it has been demonstrated that the spectral
resolution is a key factor (Worthey \& Ottviani \cite{wo97}, 
hereafter WO97, Vazdekis \cite{v99}). To take profit of a resolution higher 
than about R=2000
both the internal kinematics and the characteristics of the stellar populations must be
fitted simultaneously. Making available an extended library of stellar spectra is the
first step in this direction.

Current libraries are based on
low resolution spectroscopy (eg Jacoby et al. \cite{jaco84}, Serote Roos et al \cite{s96}, 
Pickles \cite{p98}) or they are
restricted to a limited area of the HR diagram (eg Montes et al 2000 and 
Soubiran et al \cite{s98} for F,G,K
stars). Here we present an homogeneous dataset of spectra which fills the 
deficiency of such libraries. 
The sample of stars and associated data are presented in Sect. 2. 
Sect. 3 \& 4
describe the two phases of the data processing. In Sect.
5 we evaluate the quality of the archive.
The access to the data through HYPERCAT\footnote{http://www-obs.univ-lyon1.fr/hypercat/11/spectrophotometry.html} 
or at the Centre de Donn\'ees astronomiques de Strasbourg\footnote{http://cdsweb.u-strasbg.fr/cats/III.htx}
are described in Sect. 6.

\section{Stellar content and data}

\subsection{Physical parameters of the stars}

The present sample was assembled by merging the stellar library of F,G,K stars 
(Soubiran et al \cite{s98}) serving as a basis for the TGMET program
with additional spectra taken from other observing programs and from
the near-to-line ELODIE archive (http://www.obs-hp.fr).
The spectra of the database correspond to stars which have been selected because
they have published values of \Teff or \tgm, or reliable estimations of 
the absolute magnitude \Mv, deduced from Hipparcos parallaxes. 
The main source of atmospheric
parameters is the catalogue of \FeH determinations, 1996 and 2001 versions (Cayrel de Strobel 
et al \cite{c97} and
\cite{c01}) but two other sources have been also used : 
Carney et al (\cite{c94}) and Th\'evenin (\cite{t98}). Effective
temperatures were also taken in the lists published by  Blackwell \& Lynas-Gray (\cite{b98}), 
Alonso et al (\cite{a96a}) and
Alonso et al (\cite{a99a}), or calculated from the colour indices V-K or b-y following the relations
\Teff=f(colour,\FeH) established by 
Alonso et al \cite{a96b} and \cite{a99b} for dwarfs and giants respectively. 
Each star had \tgm estimated by averaging
the determinations found in the literature, giving a half weight to old references and Str\"omgren
photometry and a double weight to \Teff determined by Blackwell \& Lynas-Gray (\cite{b98}). 
As in the TGMET library, the parameters were then labelled according to 
their reliability 
which was estimated from the
number of determinations and their standard deviation around the mean. The reliability scale runs 
from 0 to 4, with the highest reliability 4 corresponding to uncertainties lower than 80 K in \Teff, 
and 0.06 dex in \FeH. These uncertainties can reach 115 K and 0.09 dex for reliability 3, and 150 K and 
0.11 dex for the reliability 2. Reliability 1 is attributed for parameters based on old determinations or photometry, or 
presenting
large discrepancies between references. Reliability 0 means that no reference on atmospheric parameters
was found in the literature. 
The database is made of 
908 spectra corresponding to 709
different stars which are shown in the plane ($\log$ \Teff, \logg) in Fig 1 and in the 
plane ($\log$ \Teff, \FeH) in Fig. 2.  \Teff ranges from 3700 K to 13600 K, \logg ranges 
from 0.03 to 5.86 
and  \FeH ranges from -2.8 to +0.7. In these two plots, new estimations of \tgm were used instead of 
those from the literature which are incomplete. These new estimations were obtained with 
the current version
of TGMET which is still under development (a paper presenting this new version is in preparation
and a description is given in the electronic version of the archive). 
 Atmospheric parameters from the literature together with new estimations and measured line indices 
from the spectra presented here are 
given in Table 1, only available in electronic form (see sect. 6).

More information on each star is available in the header of the FITS spectra. 
It includes 
the absolute magnitude \Mv from
Hipparcos parallaxe when available. A scale of reliability was also established from the precision of the
parallaxe and V magnitude from Tycho-2 (H{\o}g et al \cite{h00}), 4 corresponding to parallaxes with a 
precision better than 10\%. Also available in the FITS headers are the spectral types, 
(V, B-V) taken from the Tycho-2
catalogue (converted from the Tycho-2 B$_\mathrm{T}$  
and V$_\mathrm{T}$) and radial velocities usually measured at the telescope. 
The on-line cross-correlation technique
was used to measure the radial velocity of strong-lined spectra, corresponding to moderate
effective temperatures up to 6500 K but for hotter stars, the radial velocity was estimated using a
least-squares deconvolution technique (Donati et al. \cite{d97}).

The database includes several stars with line profiles broadened by
rotation, macroturbulence or binarity and a few stars with spectral
peculiarities (Ap, Am, emission line stars...). These pecularities are 
indicated in the headers.

\begin{figure}[t]
\resizebox{\hsize}{!}{\includegraphics{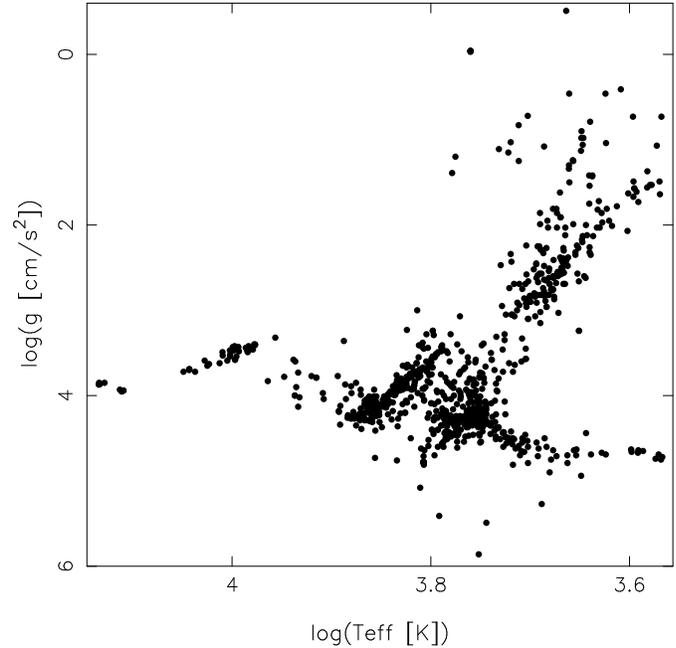}}
\caption{\Teff vs. \logg for the 709 stars of the archive. For the sake of homogeneity, 
the atmospheric parameters presented here are not from the literature which is incomplete, but were
estimated with  the current version of TGMET (Sect 5.3.3).}
\label{Teff_logg}
\end{figure}

\begin{figure}[t]
\resizebox{\hsize}{!}{\includegraphics{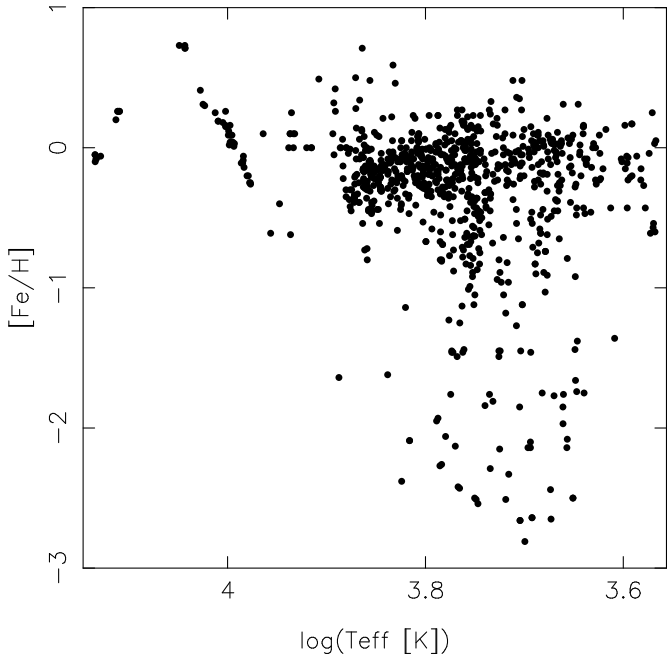}}
\caption{Distribution of the 709 stars of the archive in the plane \Teff vs \FeH
estimated with the current version of TGMET (Sect 5.3.3).}
\label{Teff_FeH}
\end{figure}

\subsection{ELODIE spectra}

ELODIE provides a spectral range of 390-680~nm 
recorded in a single
exposure as 67 orders on a 1K CCD at a mean resolving power
of 42000.
Optimal extraction and wavelength calibration are automatically performed
by the on-line reduction software TACOS. Complete information on ELODIE can be found in Baranne et al
(\cite{b96}) and in the TACOS user manual by Queloz (\cite{q96}). 
The global efficiency of the
instrument drops in the blue by a factor 5 at 440~nm for a 
solar-type star and by a factor 50 at 390~nm. Hence we decided to
limit the spectral range to $\lambda\lambda = 410 - 680 $~nm.

The typical
signal-to-noise (S/N) ratio per pixel at 550~nm is 150 (it is higher than 80 for
more than 90\% of  the spectra).
Several spectra with a lower S/N, down to 35, correspond generally to faint 
deficient stars which have been included in the database for a better sampling of the parameter space.

For the present purpose, the 908 spectra of the
database were processed to 
provide 1D spectra rebinned in wavelength and calibrated in absolute
fluxes. 
Since the main goal of the archive was to construct a library of spectra
representative of the various stellar types and an interpolated
grid of spectra covering the parameters space, all the spectra
were reduced to the rest-frame.
The archive was
prepared from the flat-fielded and order-extracted TACOS spectra 
in two main steps. (1) The orders were de-blazed, the spikes due to cosmic rays and 
telluric lines
were masked, all the
orders were connected in one spectrum, and the pseudo-continuum normalization was computed.
(2) The spectra were calibrated into ``physical'' flux, ie. above the
atmosphere.

\section{Strengthening and normalization to the pseudo-continuum}

This phase of the data processing takes the TACOS spectra
to deliver order-connected spectra with a constant wavelength step.
Both the instrumental and continuum fluxes are evaluated.

The operations follow these steps:

\begin{itemize}
\item Strengthen the spectra by dividing them by a de-blazing function
\item Compute the pseudo-continuum normalization and analyze the 
``shape'' of each order to determine a correction to the diffuse-light
subtraction.
\item Mask spikes and telluric lines.

\item Convolve and rebin with two desired resolutions (R=10000 and R=42000)
to 1D spectra.

\end{itemize}

\subsection{De-blazing and pseudo-continuum normalization}

The instrument,
hard- and soft-ware, was mostly designed as a ``velocimeter" and hence
the spectrophotometric quality was not optimized. 
The standard flat-field correction and order extraction
was applied as described in 
Baranne et al. (\cite{b96}). This procedure alone would not permit to connect the
spectra since the blaze is not accurately enough corrected, each
order presents a residual ``curvature'' of about 2\%. 
This effect is certainly due to diffuse light within the spectrograph 
which is imperfectly corrected by the TACOS program, this is a common 
difficulty with echelle spectrographs.

To solve this problem we found no other possibility but to analyze this
residual curvature on the actual spectra and correct it. Obviously, it
is a risky approach since the spectrum itself is used to determine the 
correction. To minimize possible biases, the residual curvature
is modeled with only a few parameters : each order is represented by
a second degree polynomial whose coefficients are not independent from order to 
order, but are themselves second degree polynomials of the order number.
The total effect of the diffuse light is modeled by only 3 
coefficients.

Practically, we divide the TACOS spectrum by a de-blazing function 
determined from a tungsten-lamp spectrum (internal flat-field)
and corrected using 3 metal-deficient stars (because the blaze function
appears to have slight systematic differences between internal flat-field
and star).

To analyze the ``residual'' curvature of each order, 
we normalized the spectrum to its
pseudo-continuum. The pseudo-continuum has been defined using one point
in each order and then interpolated over the whole range of wavelengths
with a $3^\mathrm{rd}$ degree spline. 
The pseudo-continuum level  
is determined by fitting a second degree polynomial
to each order. The spectrum is weighted with a mask intended to avoid as much
as possible the stellar lines, the fit is iterated after $\kappa-\sigma$ 
clipping and the remaining bias due to the weak stellar lines is statistically
corrected.
The latter correction is based of the analysis of the skewness of the 
distribution of residuals. The weighting mask is prepared from a previous
iteration of the reduction of the whole archive: for each wavelength
point it gives the probability to have a stellar line 
(we started with no weight).

Though the visual inspection of the strengthened spectra appear
satisfactory, the detection of possible biases have been a major concern
and is presented in Sect. 5.

After the correction for diffuse light was applied, the pseudo-continuum was
re-computed. These normalized spectra are \, scoped for
abundance studies, and determination of the atmospheric parameters with TGMET.

 All the information permitting calculation of the detected flux (in electrons)
is kept in the form of a table extension attached to the FITS image. The 
S/N of each pixel is also kept in a separate extension. 
The detailed description of the FITS files is given with the electronic
version of the tables and spectra.

\subsection{Cosmic rays and telluric lines}

The spikes due cosmics and defective pixels and the telluric lines have 
been masked 
following the procedure developed by Katz et al. (\cite{k98}).

\subsection{Resampling to 1D spectra}

The orders are normalized to their pseudo-continuum and
resampled to a 1D spectrum with a wavelength step of 0.005~nm using 
spline interpolation (the observations are over-sampled by about 20\%).

To produce a version of the archive at the resolution 10000
 (i.e. 30 \kms at $\lambda$ = 550~nm) the spectra are convolved 
by a Gaussian
function of FWHM = 0.054~nm, and rebinned with a step of 0.02~nm.

The high resolution version has a constant resolution of R=42000, ie.
the resolution in nm changes with the wavelength. At variance, with the
low resolution archive, the wavelength resolution, in nm, is constant 
throughout the spectra.

Note also that the convolution to produce the low resolution archive is done 
on the continuum-normalized spectra, which is not formally equivalent
to convolving the flux calibrated spectra. However the induced difference
is absolutely insignificant.

\section{Flux calibration}

The goal of the flux calibration is to determine the spectral energy
distribution above the atmosphere (the physical flux).
The standard procedure to achieve the objective (Bessell \cite{b99})
consists in determining the instrumental response and correct
for the atmospheric absorption by comparing the spectra to templates
of known spectral energy distribution observed with the same setup
between the target stars. 

Since the primary scope of the observation was not spectrophotometry,
no particular care was taken in interleaving template stars and some
observations were done through heavy atmospheric absorption. Therefore,
the standard procedure cannot be straightforwardly applied. Before entering
into the details of the procedure, we present hereafter its genesis.

The relation between the instrumental flux deduced in the previous section
and the physical flux (the flux calibration relation) primarily
reflects the effects of (1) the instrumental response (ie. the combined effect 
of the whole optics and CCD spectral sensitivity), (2) the absorption
by a clear atmosphere and (3) an additional extinction due to an atmospheric
haze.

We started from the assumptions that (1) the instrumental response is stable
or at least stable during each observing run, (2) the absorption by the
clear atmosphere can be parameterized by the airmass and (3) that
the effect of the haze can be parameterized by a {\it color excess}
defined as the difference between the measured B-V color and the
Tycho-2 B-V color.

A fit of these three functions of the wavelength using spectra of stars
with known spectral energy distribution (the details are presented below) 
resulted in a photometric
precision of about 5\% when we considered the instrumental response stable
over all observation runs.

The shape of the three functions entering the calibration relation, and the
analysis of the residuals lead to the suspicion that the correction
of the atmospheric refraction provided by the spectrograph was not fully
effective.
This was diagnosed by (1) a stronger than expected wavelength dependence of the
apparent atmospheric and haze transmissions and 
(2) a residual random ``curvature'' of the calibrated spectra 
that increases with 
airmass. Our interpretation is that the ``clear atmosphere'' function corrects
at the first order the effect of the refraction (the star is centered
on the fiber aperture by the auto-guider on the blue end of the 
atmospheric spectrum). The uncertainty on the centering of the star
(miss-guiding) results in a ``color excess'' absorbed by the haze function,
but depending on the value of the miss-centering and of the seeing a
residual curvature is left.

It was unfortunately not possible to parameterize this effect since it depends 
on a random and unknown miss-centering
and on the seeing which is not always recorded in the observing logs.
An alternative would have been to use an additional color excess to constrain
the curvature, but no such independent measurement is available for the stars
of our archive.

Adding a quadratic airmass term, another function of color excess and
allowing for variation of the instrumental response between the different
observing runs led to a photometric quality of about 3\% which was
then improved to 2\% after corrections based on internal comparisons.
We did not find hints for other effect on the photometric quality, such as 
the effect of the stress on the optical fiber or positioning of the
color correction filter that may change the instrumental response.

We will now describe the different steps of the calibrations:
(1) {\it primary} calibration based on the comparison with low spectral 
resolution flux templates (2) {\it secondary} calibration based on internal
comparison and (3) correction of the instrumental response using
template spectra with a better spectral resolution than for the
primary calibration.

\subsection{Primary calibration}

As already explained, the observations were not originally scoped for
photometry and no flux templates were explicitly observed. However,
since the archive contains bright stars, many of them are in common with
other libraries or datasets.

The largest intersection is with the collection of low resolution spectra
presented in Burnashev (\cite{b85}, CDS III/126). The spectral resolution is
about R=100 (FWHM = 5 - 6~nm), and the observations are from different 
sources which
were homogenized to a common scale. The observations were done with 
photo-electric spectro-photometers mostly a the Crimean Observatory, and
are not corrected from the interstellar extinction.

Some stars in our archive are in common with other datasets with a better
spectral resolution, but not in a number sufficient to fit the
calibration relation. For this reason we choose to first calibrate the 
archive with this dataset.
It was entered in the Hypercat Fits Archive (HFA)
with the identification L1985BURN. 

The first step was to assess the photometric quality of these templates.
The B and V magnitudes were integrated on the spectra
and compared with the measurements converted from the Tycho-2 catalogue.
We found a RMS difference on B-V of 0.05. The difference with the Tycho-2
color is used for affecting a weight to each of these template observations.

The determination of the calibration relation is done after the instrumental 
spectra from our archive are convolved to the resolution of L1985BURN.
The original reference does not precisely document this resolution which varies
with the origin of the observations, and we had to determine it. A first
guess was obtained by fitting Gaussians on strong Balmer lines and this was
fine-tuned afterwards to minimize the residuals on the strong lines. It was 
necessary to adopt slightly different resolutions for the blue and red
parts of the spectra.

The model adopted for the primary calibration relation is:

\begin{eqnarray}
\log(F_{phys}(\lambda)) &=& \log(F_{inst}(\lambda))
                        + \log(F_{inst}'(\lambda, run)) \nonumber \\ & &
                        +  a(\lambda) Airm  
                        +  b(\lambda) Airm^2 \nonumber \\ & &
                        +  c(\lambda) E_{B-V}
                        +  d(\lambda) E_{B-V}'
\end{eqnarray} 

$F_{phys}$ is the {\it physical} flux reduced above the atmosphere.
The two right-hand terms $\log(F_{inst}(\lambda)$ and $\log(F_{inst}'(\lambda, run))$
 represent the instrumental response.
The first is stable over all the observing runs while the second contains
the variable part. We separated the two terms because the second could not
be determined for all the observing runs: when the number of calibrated 
observations was less than 10 the second term was not computed and
assumed to be 0.  The run-dependent instrumental response was determined
for 9 runs out of 21, and a secondary calibration (next sub-section) was also obtained.

The next two terms depend on the airmass ($Airm$). The first of them accounts
for the clear atmosphere absorption and mean aperture effect due to
the atmospheric extinction. The second partly corrects the ``curvature''
of the spectrum due to the refraction (the length of the atmospheric
spectra is proportional to $\tan z$, $z$ being the zenithal distance).

The two last terms are parameterized by a B-V color excess, defined as the 
difference between B-V determined
on our spectra and the Tycho-2 B-V.

It is not possible to measure directly the B magnitude on our spectra since
the standard B band extend farther blue-ward than their limit. Hence it
was necessary to ``calibrate'' a different color,
by fitting a color equation as it is usually done for converting between
photometric systems. This procedure has the disadvantage that it formally
depends on the actual spectral energy distribution of the star and hence may
introduce a bias linked to the atmospheric parameters of the stars.
However, such effect may be estimated from the residuals of the fit of the
color equation, and for small color terms (ie. if our internal B is close 
to the Johnson B) it can be neglected.

Since this conversion was necessary, we used this opportunity to define 2
B bands ($b_1$ and $b_2$, converted to Johnson scale as $B_1$ and $B_2$) 
giving two estimates of the color excess. 
The difference between these two estimates gives an approximate measure of the 
residual curvature of the spectra due to the atmospheric dispersion.

\begin{eqnarray}
E(B-V) &=& (B-V)_{Tycho} - (B_1-V)_{Elodie}\\
E(B-V)'&=& B_2 - B_1
\end{eqnarray}

The $B_1$ and $B_2$ bands were respectively defined as the integral
of the flux distribution between $\lambda\lambda = 450 - 470$~nm 
and  $\lambda\lambda = 437 - 450$~nm.

The color equation was determined with the 2238 spectra of L1985BURN,
using ordinary least squares and iterative rejection of outliers, as:

\begin{eqnarray}
b_1 &=& 14.830 + 1.001 B_1 - 0.337 (B-V) \\ 
b_2 &=& 15.195 + 1.003 B_2 - 0.081 (B-V) 
\end{eqnarray}

The RMS residuals from these fits are about 0.025 mag.
The magnitudes and colors are practically measured on the instrumental 
spectra (not physical flux), and hence the measured excess is biased
due to the instrumental response and depends on the airmass. 
Though it does not change the fit to Eq. 1, in order to have independent 
functions we corrected the excess from the effect of instrumental response
and atmospheric extinction using an {\it a posteriori} fit between corrected and 
un-corrected excess.

The flux calibration relation was fitted on 369 spectra of stars in common
between our archive and L1985BURN. The different functions of the wavelength
were finally smoothed by fitting polynomials in order to erase the residual
effects of resolution miss-matching which results in spurious irregularities
in these functions (in particular around $H_\gamma$ and G-band).

\subsection{Secondary calibration}

The primary calibration  allowed us to determine the run-dependent response
for only half the runs (but more than half the spectra since the number
of stars is not equal in each run). By inter-comparing spectra of the same stars
for which several observations were available it is possible to make a {\it secondary}
calibration, using the runs where the run-dependent response was available
as ``template''.

This exercise allowed us to detect a significant change in the response
of the instrument between the observations prior to 1995 and those done after.
This change seems to be due to a modification of the differential
refraction corrector at the end of year 1994. In addition, for two other runs
the number of inter-comparisons was large enough to adopt a reliable 
run-dependent corrections. The run-dependent correction account for a RMS
photometric effect of about 0.025 mag. For 9 runs no such correction
is available at present.

\subsection{Correction of the instrumental response}

\begin{figure*}[t]
\resizebox{\hsize}{!}{\includegraphics{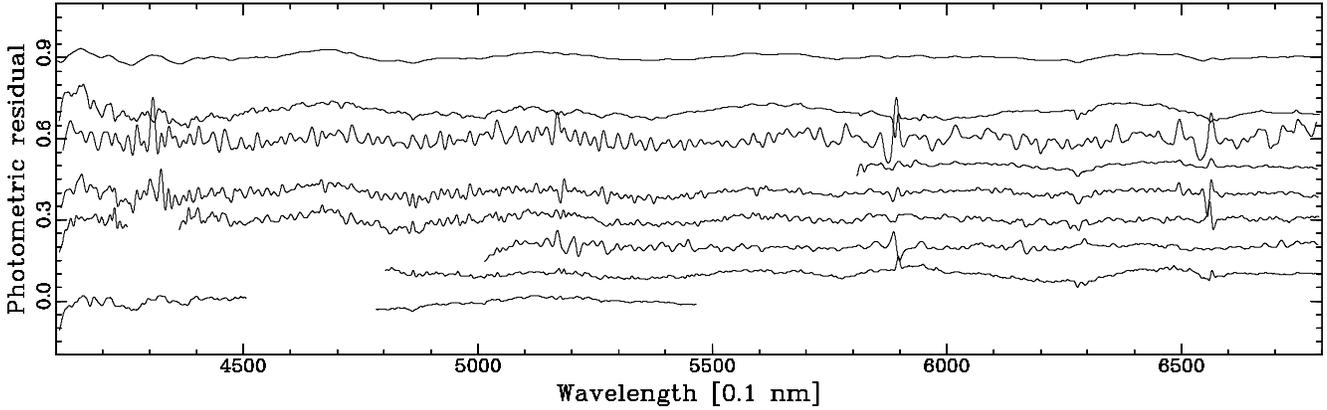}}
\caption{The mean photometric residuals of the 9 libraries used 
for the correction of the
instrumental response (before this corrections).
The ordinate is the mean difference (Elodie - reference library)
in units of relative physical flux (normalized to 1 at 555 nm).
Each dataset is shown on a different line, each being arbitrary y-shifted by
0.05. From bottom to top they are: Jones (blue and red), Serote Roos et al. 
(OHP and CFH), Jacoby, Kiehling, Danks \& Dennefeld, Gunn \& Stryker and the 
Sun. For the purpose of the display these spectra have been slightly smoothed.
The top line is the combination of all the individual comparisons which
were used to compute the correction to the instrumental response.
}
\label{fig3}
\end{figure*}

Up to this point the flux calibration can be assumed to properly correct 
the effect of atmospheric absorption and to correct at 
the best the aperture effect due to the differential refraction.
However, the instrumental response was evaluated using low-resolution spectra
and had furthermore to be smoothed. At variance with the other functions which
presumably vary smoothly with wavelength, the instrumental response may
be more chaotic.

Therefore, the last step of the calibration process consists in using
higher spectral resolution ($R \approx 1000$) to correct the instrumental
response. These reference spectra are taken from the following dataset,
referenced by their HFA identification:

\begin{itemize}

\item L96111KP1 and L96111KP2. The ELODIE database contains 191 spectra of 144 stars
from the Jones library (Leitherer et al. 1998)
($R=2800$). 
This library cover two narrow wavelength ranges, 
one in the blue part below $\lambda = 451 $~nm
the other between $\lambda \lambda= 478 - 547 $~nm. This is the highest
resolution reference for stellar spectra yet available for comparison.
This library is not corrected for the aperture effect due to differential
refraction and is not of photometric quality. After re-calibrating the 
``color'' of this spectra (defining special color-band as we did for
our spectra for the primary calibration), the mean agreement with our spectra
is satisfactory.

\item L1996SERO.  The ELODIE database has 8 spectra of 6 stars
in common with Serote Ross et al. (\cite{s96}). 
These spectra are from CFH (R=650) and OHP
(R=4000) observations. They intersect the red half of our spectral range.
The comparison with the L1996SERO spectra, after the spectral resolutions
are matched by convolving our spectra with a Gaussian, show pixel-to-pixel
variations of the order of 2\% (rms) and low-frequency variations of
typically 5\%. These figures are consistent with the comparison performed
by Serote Ross et al. (\cite{s96}) with the Silva \& Cornell (\cite{s92}) 
library.
%Our two observations of HD38858 disagree between them. ????. And none
%of our observations is consistent with L1996SERO/00022 .????.

\item L1984JACO. We have 2 spectra for 2 stars (HD094028 \& HD028099)
in common with the Jacoby library (R=1200) 
which cover the full wavelength range.
%% On pourrait recuperer deux autres etoiles dans l'archive Elodie de l'OHP:
%% HD260655 n19881025 #15
%% HD027685 n19951213 #24
The two spectra compared with L1984JACO reveal a significant gradient
(the difference over the whole range of wavelength are
respectively 29\% and 14\% for the first and second spectrum)
and a pixel-to-pixel variation of 0.5\% rms.
The B-V color integrated from the L1984JACO spectra disagrees with 
the Tycho-2 value in amounts that explain the differences with our spectra.

\item L1987KIEH. We have 13 spectra of 7
stars from the dataset of Kiehling (1987) (R=800).

\item L1994DANK. We
have 17 spectra from 14 stars in common with 
Danks \& Dennefeld (\cite{danks}) (R=1500). The red half of our wavelength 
range intersects with L1994DANK.

\item L1983GUNN. The  Gunn \& Stryker (\cite{gunn}) library (R=500) counts 5
stars in common with our archive.

\item L1984NSOA. Finally we compared our 7 solar spectra with the 
{\it Solar flux atlas} (Kurucz et al. \cite{k84}) 
(R=500000, also stored in HFA with R=10000).

\end{itemize}

We made all the individual comparisons between our spectra, 
convolved by a Gaussian to match the resolution, and these references.
Because the broad-band variations are probably due either
to random photometric errors in our archive (certainly not systematic), 
correction of the interstellar extinction (it was not always possible
to de-correct the templates) or errors in the templates,
we subtracted a low-degree polynomial (1 to 3 depending for which dataset)
to these comparisons. 

The individual 
comparisons (see Fig.~\ref{fig3}) were then combined and produced a mean 
residual convolved
to a resolution of 2 nm FWHM (each order of the original spectrum covers about
4 nm and no correction of the instrumental response at a lower scale is 
desired). This was used to correct the instrumental  response.

\begin{figure*}[t]
\resizebox{\hsize}{!}{\includegraphics{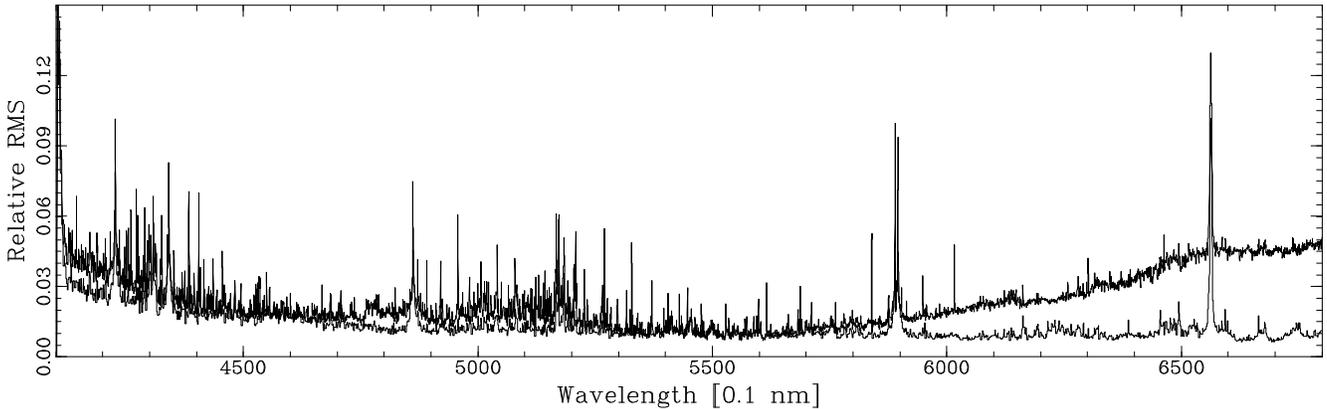}}
\caption{Photometric precision deduced from the comparison of multiple
observations of the same star.
The ordinate are the rms of the 358 pair comparisons (ln(Flux$_2$/Flux$_1$))
(the spiky appearance is due to the logarithmic comparison).
The higher curve the the total RMS while the lower one is after third
degree polynomials were subtracted from each individual comparison: This 
disentangles the effect of photon noise and uncertainty on the long range
flux calibration (due to the differential refraction).
To obtain the order of magnitude of the photometric error on single
observation, the ordinates should be divided by $\sqrt{2}$.
}
\label{fig4}
\end{figure*}

We also made an unsuccessful attempt to compare our \, archive with the Pickles
(\cite{p98}) library (L1998PICK): all the spectra were compared to the nearest 
(in the \Teff - \logg - \FeH \, space) template in L1998PICK. But the mapping of
the parameters space in L1998PICK is not dense enough and the residuals
are too affected by the distance in the parameter space to permit a useful
analysis. We did not attempt to interpolate the spectra within the Pickles
library.

\section{Quality control}

To check the reliability of our data and reduction we used the 
following tests: (1) analyze the inter-comparison of repeated observations
and analyze internal consistency,
(2) compare fully reduced spectra with other libraries and 
(3) compare Lick indices for stars with published measurements.

There are a priori three possible sources of discrepancy to search:
(1) errors on the instrumental response and/or correction of atmospheric
extinction are likely to result in low and intermediate 
spatial frequencies variations.
(2) inaccurate subtraction of the diffuse light may be diagnosed as
significant mismatch of the strongest lines and (3) inaccurate correction
for the blaze will lead to oscillations with a one-order period.

\subsection{Internal consistency}

\subsubsection{Multiple observations of the same star}

There are in total 358 repeated observations for 104 stars in the archive.
After giving a double weight to the observations in the runs for
which the run-dependent response was computed from primary calibration
and half weight to the observations done with an airmass larger than 1.5
(their photometric quality is lower because of the differential refraction),
we find a global RMS dispersion for the pairwise comparison of about 3.5\%,
see Fig.~\ref{fig4}. Without weighting, the dispersion is 4\% and the difference 
was a hint for the lowest 
photometric quality of high airmass observations, hence leading to suspicion 
about the refraction corrector. Since it is a pairwise comparison, 
the mean photometric error on individual observations is: 
$3.5 / \sqrt{2} = 2.5\%$. (Note that this error does not account for 
possible systematic errors affecting uniformly the whole archive, it will be
discussed in Sect. 5.2 \& 5.3).

Most of the photometric errors lie in residual ``curvature'' of the spectra.
After subtracting a third degree polynomial to each pairwyse comparison, 
the RMS dispersion becomes 0.7\%, corresponding to a mean photometric 
precision of 0.5\%, in full agreement with the mean S/N ratio. No hint for
a modulation at the scale of one order of the echelle spectra is found.

\subsubsection{Modeled spectra}

Random errors on the strengthening of the spectra are probably in average
smaller than 0.5\% because otherwise they would increase the dispersion of
the pairwyse internal comparison (after the third degree polynomial is 
subtracted). 

The new version of the TGMET program that will be described in a separate
paper allowed us to generate interpolated spectra for a given position
in the \tgm space. The difference between the spectra in the archive and these
modeled spectra can potentially uncover possible random errors on the
strengthening.

Therefore we computed and averaged the differences to the modeled 
spectra for all the 
spectra in the archive. 
The mean residuals
were searched for a modulation at the scale of one order,
either in the spectrum or in its Fourier transform. We did not detect
any signal. We made simulations that told that a 0.5\% (RMS) modulation 
would  probably have been detected, but a 0.2\% modulation would not.

\subsection{External comparisons}

The comparisons between our physically calibrated spectra and other libraries
were used in the previous section to compute the correction to the 
instrumental response. The goal was to determine the variation of the response
at the scale of 1 to 5 nm that was not constrained by the primary calibration.
Therefore, by construction, the mean external comparison is flat
(ie. may only varies smoothly with the wavelength). A careful analysis of the
individual comparison could assess the level of stability of the 
instrumental response.

The patterns in the correction to the instrumental response were consistently
found in the different individual comparisons and no significant effect
was found after this correction was applied. However, the scarcity of the
available comparisons did not allow us to measure a photometric precision.

\subsection{Comparison of Lick indices}

The Lick indices (see eg. WO97) measure equivalent
widths of features defined in bands of 2 to 5~nm (hence of the width of
about 1/2 or 1 order of the echelle spectra). Their measurements are
then sensitive to both incorrect strengthening of the spectra and errors
in the instrumental response.

The spectral range of the Elodie archive allowed us to measure most of the Lick
indices. Only the blue CN indices are missed as well as the H$\delta$
indices (H$\delta_A$ and H$\delta_F$) defined in WO97.

We used the definitions from WO97 or Cardiel et al. (\cite{c98}) to measure the
indices for all the spectra in the archive. The results are reported in
Table 1 given in the electronic version only.

The measurement errors computed from the photon noise 
(see eg. Prugniel et al  \cite{p01}) is in the present case negligible because
of the extremely high S/N per \AA. The main source of error is due to the
photometric calibration and may be systematic. The comparison the Lick 
indices measured in WO97 constrain the magnitude of these systematic errors.

The archive has 102 stars which have been previously measured in WO97. 
The comparisons are reported in Table 2.

\setcounter{table}{1}
\begin{table}[ht]
\caption[]{Comparison of Lick indices:
Col. 1: Name of the Lick index
Col. 2: Number of spectra used for the comparison
Col. 3: Mean value of the index
Col. 4: Mean shift (Elodie) - (WO97)
Col. 5: RMS dispersion of the comparison
Col. 6: Majorant of photometric error
}
\begin{flushleft}
\begin{tabular}{| l | r | r | r | r | r |} \hline
Index & \# & Mean & Shift & RMS & Error\\
\hline

Ca4227      & 101  &   0.689  &	 -0.091  &   0.278 &  -0.0073\\
G4300       & 102  &   3.858  &	  0.083  &   0.464 &   0.0024\\
H$\gamma_A$ & 102  &  -1.765  &	  0.392  &   0.527 &   0.0089\\
H$\gamma_F$ &  99  &   0.678  &	  0.177  &   0.337 &   0.0084\\
Fe4383      & 102  &   2.715  &	 -0.341  &   0.692 &  -0.0067\\
Ca4455      & 101  &   0.723  &	 -0.267  &   0.250 &  -0.0118\\
Fe4531      & 101  &   2.299  &	 -0.078  &   0.335 &  -0.0017\\
Fe4668      & 102  &   2.214  &	 -0.183  &   0.712 &  -0.0021\\
H$_\beta$   & 103  &   2.706  &	  0.061  &   0.192 &   0.0021\\
Fe5015      & 102  &   3.393  &	 -0.197  &   0.449 &  -0.0026\\
Mg$_1$      & 103  &   0.040  &	  0.003  &   0.011 &   0.0030\\
Mg$_2$      & 103  &   0.125  &	 -0.005  &   0.011 &  -0.0054\\
Mg$_b$      & 100  &   2.335  &	  0.117  &   0.247 &   0.0036\\
Fe5270      & 100  &   1.777  &	 -0.088  &   0.254 &  -0.0022\\
Fe5335      & 102  &   1.515  &	 -0.035  &   0.279 &  -0.0009\\
Fe5406      & 103  &   0.895  &	  0.076  &   0.279 &   0.0028\\
Fe5709      & 101  &   0.505  &	 -0.046  &   0.187 &  -0.0020\\
Fe5782      & 101  &   0.313  &	 -0.095  &   0.168 &  -0.0047\\
NaD         & 103  &   1.415  &	 -0.165  &   0.408 &  -0.0051\\
TiO$_1$     & 103  &   0.006  &	 -0.006  &   0.011 &  -0.0057\\
TiO$_2$     & 103  &   0.011  &	  0.005  &   0.009 &   0.0054\\

\hline
\end{tabular}
\end{flushleft}
\end{table}

We do not find systematic effects that would result from errors in the
instrumental errors and the residual dispersions are compatible with
the errors in the WO97 measurements (the internal error due to the
photon noise in the archive is totally negligible).

\begin{figure}[t]
\resizebox{\hsize}{!}{\includegraphics{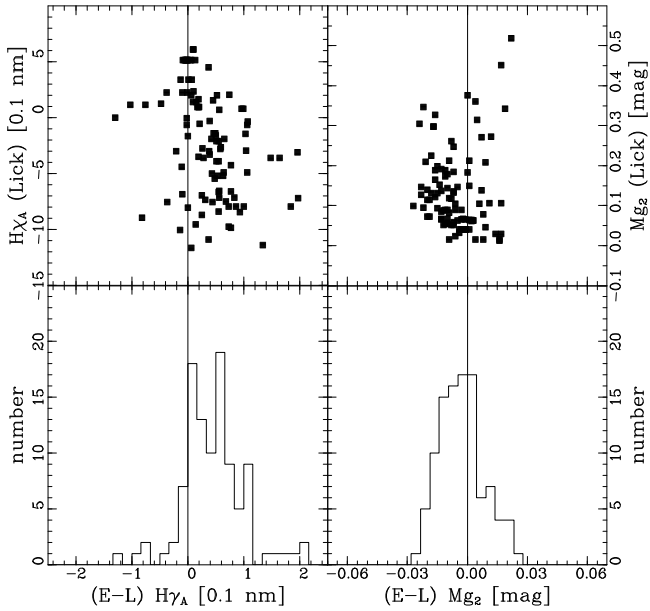}}
\caption{Comparison of Lick indices Mg$_2$ and H$\gamma_A$ with WO97.
The abcissae are the difference (Elodie - WO97) between the Lick indices.
The top panels show the distribution as a function of the value in WO97,
and the bottom panels the histograms of the differences.
}
\label{fig5}
\end{figure}

The systematic shifts are always smaller than the dispersion except for
Ca4455 (where it is equal) and for Fe4383 the shift may be significant.
However, we cannot attribute this to an error in our measurements and in 
particular we note that the shift for Fe4383 is similar to the one found by 
WO97 when they compared the original Lick measurements with the spectra
of the Jones library (note that our flux calibration is not fully
independent from the Jones library since this latter was used among others
to correct the instrumental response). The Ca4455 index is not in
the range of the Jones spectra and we cannot comment of the shift we found.
Fig.~\ref{fig5} presents the distributions of the differences (us - WO97) for 
two indices: Mg$_2$ and H$\gamma_A$. The first one is very important for 
extragalactic studies, it is defined on a broad region (53~nm between the
extrema of the red and blue side-bands). Though it is potentially sensitive 
to the uncertainties on the flux calibration described above, the agreement
between us and WO97 is excellent. The second index presented in Fig. 5,
H$\gamma_A$, is narrower (14~nm) but it is located in the blue region of 
our spectra were our calibration is probably less accurate. There
is a systematic shift with respect to WO97, but the rms residual is
fully accounted for by the errors on WO97.

\section{Data format and public access}

The spectra are stored in HFA in the form of FITS files
containing 4 extensions (Gr{\o}sbol et al \cite{g88}):

The first extension is a 1D image header data unit (HDU) containing
the spectra calibrated in wavelength expressed in the rest-frame
of each star and sampled between 410 and 680~nm. The high resolution 
version is normalized with a step between the sampling points of 0.005~nm
and the low resolution version is given in physical flux with a step of
0.020~nm.

Another image HDU (EXTNAME=NOISE), with the same sampling,  
contains the noise attached to the values in 
the primary HDU (in the unit of this spectra). Pixels are independent.

The other extensions are the flux calibration relations which enable to
convert the instrumental flux into physical or normalized
flux. They are binary table extensions and the relations are coded
as spline representations. The details are given in the Pleinpot user's
manual\footnote{http://www-obs.univ-lyon1.fr/$\sim$prugniel/cgi-bin/pleinpot/pleinpot.html},
and the conversions can be done on-the-fly with the pipeline attached to 
HFA.

The physical flux calibration (EXTNAME=FCAPHY) and the normalized flux 
relation (EXTNAME=FCANOR) have a spectral resolution of about 2~nm.

Each header, which includes the keywords created during the observations
and TACOS reduction, 
have been completed with useful information on the star : 
spectral type, Tycho-2 V and B-V, absolute magnitude \Mv and 
\tgm with their reliability flag,  and \vr. The nominal FWHM of the spectrum 
can also be retrieved in the header.

In addition to the spectra of the stars, the archive contains the different
intermediate files used along the data processing. They are described in the
electronic version of the paper.

\section{Conclusion}

We have presented an archive of 908 spectra of 709 stars at two resolutions,
R=10000 and R=42000, and in the wavelength range 
$\lambda\lambda = 410 - 680 $~nm. The {\it high resolution} version is archived
normalized to its pseudo-continuum while the {\it low resolution} is
given in physical flux (above the earth atmosphere) normalized to one
at $\lambda = 555$~nm.

The mean S/N per pixel in the high resolution version of the archive is
150 (S/N $\approx$ 700 per 0.1 nm), the photometric precision of the 
strengthening of the orders of the
echelle spectra is better than 0.5~\%, the precision in bands of about 
10-20~nm is 0.5~\% and the overall (broad-band) precision is 2.5~\%, limited by
the effect of differential refraction.

The stars in this archive cover a large range of the parameter's space, 
though the mapping is still crude for hot and very cool stars.

This archive supersedes the TGMET (\cite{s96}) library and is being used
for automatic determination of the stellar parameters \tgm by direct
comparison of spectra with interpolations in the archive. The new version
of the TGMET program is still under development and will enableto determine
the parameters from any R$>$10000 spectra, independently of the detailed 
observational setup. It will be presented in a future paper. 

The archive will also be used to synthesize extragalactic stellar populations 
and to use these high resolution synthetic spectra as kinematical templates
to determine simultaneously the parameters of the stellar population and
the line-of-sight velocity distribution.

The archive itself will be completed with new observations and spectra
from the OHP archive with the goal of improving the mapping of the 
parameters space and the quality of the photometric calibration.

\begin{acknowledgements}
We thank the director and technical staff of Observatoire de Haute-Provence
for their constant support of our observing programs and for providing
some spectra retrieved from the ELODIE archive.
We thank all the observers who provided their observations to complete 
the database, especially the COROT team :
D. Ballereau,
J.-C. Bouret,
C. Catala,
T. Hua,
D. Katz, F. Lignieres,
T. Lueftinger and
C. Van't Veer.
We acknowlegde, with gratitude, financial support from the GDR Galaxies, CNRS,
France. 
This work made extensive use 
of the SIMBAD and VIZIER databases, operated at CDS, Strasbourg, France.

\end{acknowledgements}


\begin{thebibliography}{}

\bibitem[1996a]{a96a} Alonso A., Arribas S., Mart\'\i nez-Roger C., 1996a, A\&A 313, 873

\bibitem[1996b]{a96b} Alonso A., Arribas S., Mart\'\i nez-Roger C., 1996b, A\&AS 117, 227

\bibitem[1999a]{a99a} Alonso A., Arribas S., Mart\'\i nez-Roger C., 1999a, A\&AS 139, 335

\bibitem[1999b]{a99b} Alonso A., Arribas S., Mart\'\i nez-Roger C., 1999b, A\&AS 140, 261

\bibitem[1996]{b96} Baranne A., Queloz D., Mayor M., Adrianzyk G., Knispel G.,
Kohler D., Lacroix D., Meunier J.-P., 1996, A\&AS 119, 373

\bibitem[1999]{b99} Bessell M.S., 1999, PASP 111, 1426. 

\bibitem[1985]{b85} Burnashev V. I., 1985, Abastumanskaya Astrofiz. Obs., Byull., 59, 83 (CDS III/126)

\bibitem[1998]{b98} Blackwell D.E., Lynas-Gray A.E., 1998, A\&AS 129, 505

\bibitem[1998]{c98}  Cardiel N., Gorgas J., Cenarro J., Gonzalez J., 1998, A\&AS 127, 597

\bibitem[1994]{c94}  Carney B.W., Latham D.W., Laird J.B., Aguilar L.A., 1994, AJ 107, 2240.

\bibitem[1997]{c97} Cayrel de Strobel G., Soubiran C., Friel E.D., Ralite N., Fran\c cois P.,
1997, A\&AS 124,299

\bibitem[2001]{c01} Cayrel de Strobel G., Soubiran C.,  Ralite N., 
2001, A\&A, submitted.

\bibitem[1994]{danks} Danks A.C., Dennefeld M., 1994, PASP 106, 382

\bibitem[1997]{d97} Donati J.-F., Semel M., Carter B.D., Rees D.E., Cameron A.C.,
1997, MNRAS 291, 658

\bibitem[1997]{f97} Fioc M., Rocca-Volmerange B., 1997, A\&A 326, 950.

\bibitem[1988]{g88} Gr{\o}sbol P., Harten R. H., Greisen E. W., Wells D. C.,
1988, A\&AS 73, 359

\bibitem[1983]{gunn} Gunn J.E., Stryker L.L., 1983, ApJS 52, 121

\bibitem[2000]{h00}  H{\o}g E., Fabricius C., Makarov V.V., Urban S., Corbin T.,
    Wycoff G., Bastian U., Schwekendiek P., Wicenec A., 2000, A\&A 355, 27.

\bibitem[1984]{jaco84} Jacoby G.H., Hunter D.A., Christian C.A., 1984, ApJS 56, 257

\bibitem[1998]{k98} Katz D., Soubiran C., Cayrel R., Adda M., Cautain R., 1998,
A\&A 338, 151

\bibitem[1987]{k87} Kiehling R., 1987, A\&AS 69, 465

\bibitem[1984]{k84} Kurucz R.L., Furenlid, I., Brault, J., Testerman L., 1984,
\\(ftp://ftp.noao.edu/fts/fluxatl/)

\bibitem[1998]{p98} Pickles A. J., 1998, PASP 110, 863

\bibitem[2001]{p01} Prugniel Ph., Maubon G., Simien F., 2001, A\&A in press

\bibitem[1996]{q96} Queloz D., 1996, ELODIE user's guide, http://www.obs-hp.fr/

\bibitem[1996]{s96} Serote Roos M., Boisson C., Joly M., 1996, A\&AS 117, 93 

\bibitem[1992]{s92} Silva D.R., Cornell M.E. 1992,  1992, ApJS 81, 865

\bibitem[1998]{s98} Soubiran C., Katz D., Cayrel R., 1998, A\&AS 133, 221

\bibitem[1998]{t98} Th\'evenin F., 1998, BICDS 49

\bibitem[1999]{v99} Vazdekis A., 1999, ApJ 513, 224

\bibitem[1997]{wo97} Worthey G., Ottaviani D. L., 1997 ApJS 111, 377
\\(http://astro.sau.edu/~worthey/html/system.html) (WO97)

\end{thebibliography}
\end{document}